\documentclass[aps,pra,twocolumn,superscriptaddress]{revtex4-2}
\usepackage{bbm}
\usepackage{mathrsfs}
\usepackage{amsmath}
\usepackage{amsfonts}
\usepackage[colorlinks=true,linkcolor=blue,urlcolor=blue,citecolor=blue,anchorcolor=blue]{hyperref}
\usepackage{graphicx,epstopdf}
\usepackage{subfigure}
\usepackage{epsfig}
\usepackage{dcolumn}
\usepackage{bm}
\usepackage{color}
\usepackage{natbib}
\usepackage{amssymb}
\usepackage{xcolor}
\usepackage{braket}
\usepackage{ulem}
\usepackage{float}
\usepackage{lipsum}
\usepackage{pifont}
\usepackage{titlesec}

\definecolor{newtxtcolor1}{rgb}{0.8, 0, 0.2}
\definecolor{newtxtcolor2}{rgb}{0.8, 0, 0}

\begin{document}

\title{Enhanced Charging in Multi-Battery Systems by Nonreciprocity}

\author{Hua-Wei Zhao}
\affiliation{School of Physics, Beihang University,100191,Beijing, China}

\author{Yong Xie}
\affiliation{School of Physics, Beihang University,100191,Beijing, China}

\author{Xinyao Huang}
\email[Corresponding author: ]{xinyaohuang@buaa.edu.cn}
\affiliation{School of Physics, Beihang University,100191,Beijing, China}

\author{Guo-Feng Zhang}
\email[Corresponding author: ]{gf1978zhang@buaa.edu.cn}
\affiliation{School of Physics, Beihang University,100191,Beijing, China}
\begin{abstract}
Quantum batteries (QBs), harnessing quantum systems to transfer and store energy, have garnered substantial attention recently, enabling potentials in enhanced charging capacity, increased charging power, and device miniaturization. 
However, constrained by the weak interaction between the quantum nodes, the implementations of QB networks exhibit limited charging performance.
In this work, we propose an efficient approach to improving charging in multi-battery systems by capitalizing on nonreciprocity. By constructing non-Hermitian Aharonov-Bohm triangles to establish unidirectional energy transfer in both cascaded and parallel configurations, we can achieve a significant enhancement of the stored energy in QBs especially in the weak interaction regime.
Remarkably, the nonreciprocal cascaded setups display an exponentially increasing gain in the battery energy as the charging distance lengthens compared to the reciprocal counterparts. Furthermore, we demonstrate that nonreciprocity can also lead to the same enhancement in the charging power of QBs, accelerating the charging processes.  Our findings provide a practical pathway for enhancing the charging performance of QBs and exhibit the potentials for constructing efficient QB networks.
\end{abstract}
\maketitle
\textit{Introduction.---}
In contrast to traditional electrochemical batteries, quantum batteries (QBs) capitalize on quantum features to enable energy storage and transfer within quantum devices~\cite{RevModPhys.96.031001}. They carry out the charging process through the utilization of direct interactions between the charger and the QBs, which presents a promising avenue for implementing more efficient energy storage devices with greater energy storage, higher charging power and smaller size~\cite{Ito2020CollectivelyEH,PhysRevLett.118.150601,PhysRevLett.120.117702,Shastri_2025}. To date, various theoretical models of QBs have been proposed such as Dicke model battery~\cite{PhysRevLett.120.117702,Quach2020SuperabsorptionIA}, spin chain battery~\cite{PhysRevA.97.022106,PhysRevA.106.032212} and strongly interacting
Sachdev-Ye-Kitaev fermionic battery~\cite{PhysRevLett.125.236402}, demonstrating the superiority of QBs over their classical counterparts in charging processes. 
However, constrained by the fragility of quantum properties, the practical implementation of QBs is still at an initial stage. A outstanding challenge lies in the weak interaction strength between quantum systems, which restricts the charging efficiency and distance of the QBs~\cite{PhysRevLett.122.047702,PhysRevLett.122.210601,PhysRevE.102.052109,PhysRevB.104.245418,PhysRevLett.131.240401,RevModPhys.96.031001,PhysRevLett.132.090401}. This is due to the fact that the implementation of strong interaction always requires stringent experimental conditions, and strong-interacting systems are prone to instability because of their elevated sensitivity to environmental perturbations~\cite{PhysRevLett.128.140501,PhysRevResearch.2.023113}.

As a fundamentally distinct way of energy transfer between physical systems, nonreciprocity, stemming from the violation of the Lorentz reciprocity theorem, endows the system with the capacity to exhibit disparate responses for forward and backward energy transmission~\cite{RevModPhys.29.651,RevModPhys.21.463,PhysRevB.30.3277}. Beyond its functional applications in devices for unidirectional signal routing~\cite{Bi2011OnchipOI,2014isonatphy,Kang2011ReconfigurableLO,Kang2011ReconfigurableLO,PhysRevX.4.021019,PhysRevApplied.4.034002,PhysRevX.7.011007,Jing2022QuantumSP,PhysRevX.5.021025,PhysRevX.3.031001,PhysRevX.5.041020,PhysRevApplied.7.024028,PhysRevA.97.053812,PhysRevLett.133.136601}, nonreciprocity also provides an innovative means of manipulating captivating phenomena, including exceptional topology~\cite{PhysRevLett.124.250402,PhysRevApplied.15.044041,PhysRevResearch.6.013031,PhysRevB.110.045122,wang2023chiralexcitationflowsmultinode}, phase transitions~\cite{Fruchart2021NonreciprocalPT,PhysRevLett.132.127401,PhysRevLett.132.193602} and quantum correlations~\cite{PhysRevApplied.22.064001,PhysRevLett.132.120401,PhysRevLett.133.043601,PRXQuantum.4.020344,PhysRevLett.125.143605}.
Recent studies have further broadened the exploration of nonreciprocity in manipulating energy transfer between quantum systems~\cite{PhysRevLett.132.210402,PhysRevApplied.23.024010}. 
However, the investigation into the impact of nonreciprocity on the charging performance of quantum multi-battery models remains largely unexplored.

\begin{figure}
    \centering
    \includegraphics[angle=0,width=0.9\linewidth]{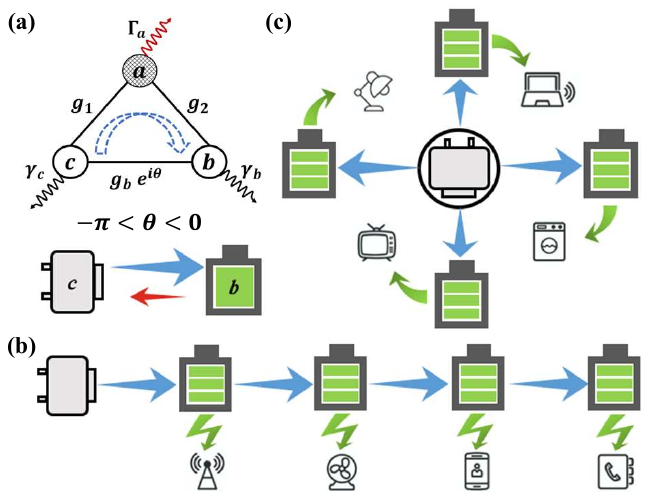}
    \caption{(a) A non-Hermitian AB triangle is composed of a QC mode $c$ with dissipation rate $\gamma_{c}$ and a QB mode $b$ with dissipation rate $\gamma_{b}$, both of which are coupled to the intermediate mode $a$ with dissipation rate $\Gamma_{a}$. $g_1$ and $g_2$ represent the corresponding coupling strength, respectively. The strength and phase of the direct interaction between $b$ and $c$ are denoted as $g_b$ and $\theta$. Nonreciprocal energy flow from $c$ to $b$ can be achieved by tuning $\theta\in\{-\pi,0\}$. (b) Cascaded charging: $N$ batteries are connected as a chain. Each mode interacts solely with its nearest-neighbor modes for transferring energy. (c) Parallel charging: $N$ batteries connect to a common charger $c$ for charging simultaneously.
    }
    \label{Model}
\end{figure}

In this study, we demonstrate the enhanced charging capabilities brought about by nonreciprocity within multi-battery systems featuring both cascaded and parallel configurations. The nonreciprocal energy transfer is achieved by introducing lossy intermediate modes to construct non-Hermitian Aharonov-Bohm (AB) structures~\cite{PhysRevLett.81.5888,PhysRevA.107.023703,Huang2021LossinducedN,PhysRevLett.124.070402,PhysRevLett.129.220403,PhysRevResearch.4.L032046}. 
The intermediate modes can be implemented in a wider range of classical (or quantum) systems, such as optical cavities~\cite{PhysRevA.97.053812},
waveguides~\cite{PhysRevA.96.053845}, micromechanical oscillators~\cite{PhysRevApplied.22.034060,PhysRevA.108.033515,PhysRevX.7.031001} and atomic ensembles~\cite{PhysRevLett.126.223603}. 
As illustrated in Fig.~\ref{Model}(a), the lossy mode $a$ establishes an additional indirect transmission channel between the quantum charger (QC) and the QB. The non-Hermitian AB triangle is implemented by adjusting the synthetic magnetic flux within the loop through modulating the phase ($\theta$) of the QC-QB interaction. When $\theta\in\{-\pi,0\}$, destructive interference between direct and indirect channels for backward energy transfer enables forward energy transmission to suppress backward energy reflux, leading to nonreciprocal charging from the QC to the battery~\cite{supp}. 

By incorporating the nonreciprocal charging unit into multi-battery systems, we can achieve a substantial improvement of the stored energy within QBs, particularly in the weak interaction regime. In contrast to their reciprocal counterparts, cascaded setups present an exponentially growing gain in battery energy as the charging distance expands. Meanwhile, in parallel systems where QBs are charged by a common charger, an energy gain improvement of up to a factor of 4 can be observed. As an additional crucial indicator for quantifying charging performance, we further demonstrate that, when maximizing the charging power of QBs in cascaded and parallel setups, respectively, the same gain factor relative to energy storage can be attained. These findings deepen our understanding of the advantages of nonreciprocity in energy storage and transfer, offering promising platforms for controlling energy flow within quantum networks.

\textit{Cascaded nonreciprocal charging.---}
As illustrated in Fig.~\ref{Model}(b), we consider the QC and a series of QBs are linearly interacted in cascade. The QC is also driven by a continuous optical field with the amplitude $\xi$, which provides a stable power source for its operation. The Hamiltonian of the cascaded charging setup in the reference frame rotating at the driving field frequency $\omega_{l}$ can be given as ($\hbar=1$)~\cite{supp}:
\begin{equation}
 \begin{aligned}
&H_{\mathrm{c}}=g_b^{(1)} e^{i \theta_1} c b_1^\dagger + \sum_{k = 1}^{N-1} g_b^{(k+1)} e^{i \theta_{k+1}} b_k b_{k+1}^\dagger+\xi c\\
&+g_1^{(1)} c a^\dagger_1+ \sum_{k = 2}^{N} g_1^{(k)} b_{k-1} a^\dagger_{k}+ \sum_{k = 1}^N g_2^{(k)} a_k b_k^\dagger+H.c.,
 \end{aligned}
 \label{cascadedHam}
\end{equation}
where the QC and QBs are modeled as the bosonic modes $c$ and $b_{k}, k\in\{1,2,..,N\}$, respectively. $g_b^{(k)}$ and $\theta_k$ represent the strength and phase of the nearest-neighbor linear interaction in the cascaded charging setup, respectively. The intermediate modes $a_{k}$ are added to achieve nonreciprocal energy exchange between the nearest-neighbor system modes. We assume that all modes are resonant with respect to the driving field for simplicity.

Starting with the chain without intermediate modes (denoted as reciprocal case I), the energies of the batteries can be derived by solving the corresponding Heisenberg equations. For the terminal battery $b_{N}$, i.e., the remotest QB from the QC, the stored energy in steady state is given as:
\begin{equation}\label{Ebrn}
    \begin{split}
    &E^{\mathrm{r1}}_{N}/\omega= 
[\frac{2^{N+1}g_b^{N}\xi}{\sum^{(N+1)/2}_{j=0}a(j)g_b^{N+1-2j}\gamma^{2j}}]^2 \  N=1,3\ldots\\
    &E^{\mathrm{\mathrm{r1}}}_{N}/\omega=
    [\frac{2^{N+1}g_b^{N-1}\xi}{\sum^{N/2}_{j=0}b(j)g_b^{N-2j}\gamma^{2j}}\frac{g_b}{\gamma}]^2\quad \quad N=2,4\ldots
    \end{split}
\end{equation}
where $g_b =g_b^{(k)}$ and $\gamma_c=\gamma_b^{(k)}=\gamma$ is assumed for simplicity with $\gamma_c$ and $\gamma_b^{(k)}$ being the dissipation rate of $c$ and $b_{k}$, respectively. $a(j)$ and $b(j)$ are constants~\cite{supp}. It is obvious that the amount of energy charged into the terminal battery $b_{N}$ differs in reciprocal case I[Eq.~\eqref{Ebrn}], which is determined by the parity of the charging system (whether the total number $N$ of batteries is even or odd). 

\begin{figure}
    \centering
    \includegraphics[angle=0,width=\linewidth]{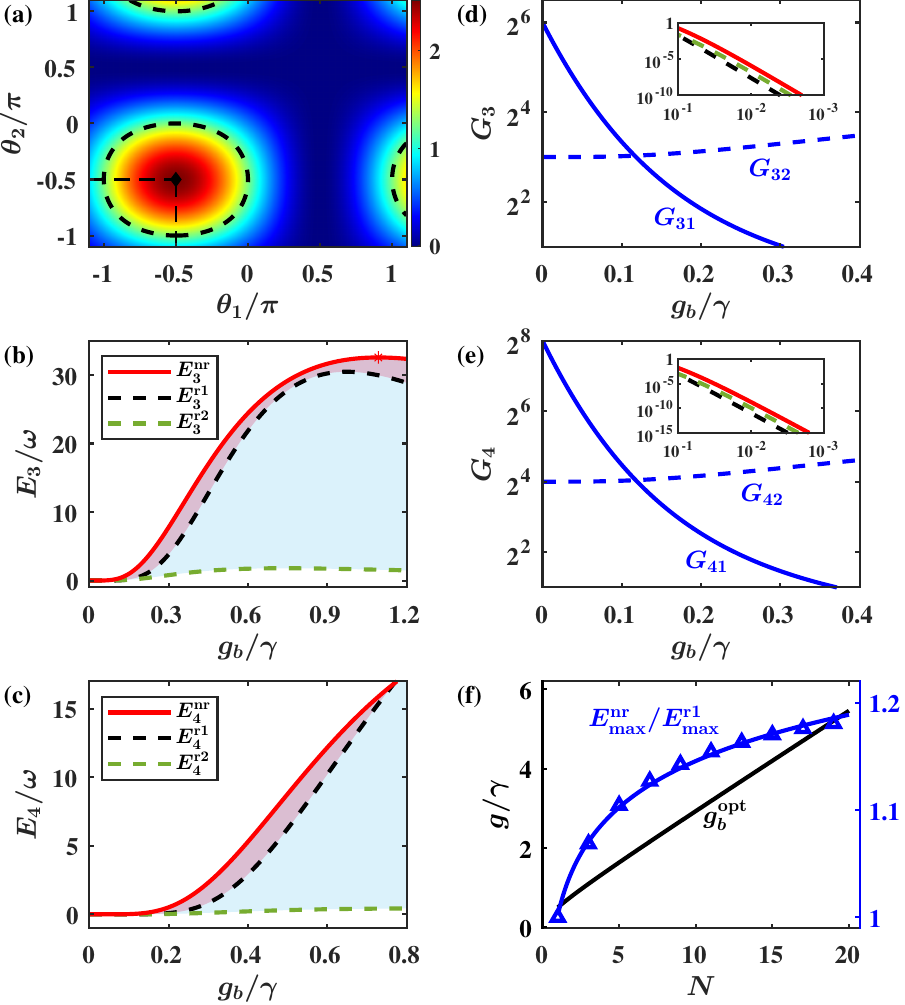}
    \caption{(a) $E_{2}/\omega$ as the function of $\theta_1$ and $\theta_2$ under $g_b/\gamma=0.1$ when $N=2$.  (b) Steady-state $E_3^{\mathrm{nr}}/\omega$, $E_3^{\mathrm{r1}}/\omega$ and $E_3^{\mathrm{r2}}/\omega$ as functions of $g_b/\gamma$ when $N=3$. (c) Steady-state $E_4^{\mathrm{nr}}/\omega$, $E_4^{\mathrm{r1}}/\omega$ and $E_4^{\mathrm{r2}}/\omega$ as functions of $g_b/\gamma$ when $N=4$. (d) The gain factor $G_{31}$ and $G_{32}$ as functions of $g_b/\gamma$ when $N=3$. (e) $G_{41}$ and $G_{42}$ as functions of $g_b/\gamma$ when $N=4$. The two insets plot the corresponding steady-state energy of the terminal battery in three cases. (f) The optimal interaction strength $g_{b}^{\mathrm{opt}}$ of nonreciprocal charging and the ratio  $E_{\max}^{\mathrm{nr}}/E_{\max}^{\mathrm{r1}}$ when $N$ belongs to odd number. The left parameters are fixed as $\Gamma = \gamma_c=\gamma_b^{(k)}=0.1\omega$, $g_1^{(k)}=g_2^{(k)}=\sqrt{g_b\Gamma/2}$ and $\xi=\omega$.}
  \label{Fig2}  
\end{figure}

To investigate the performance of nonreciprocal cascaded charging, the coupling strengths between the intermediate modes and the charging modes should firstly be tuned to optimize the steady-state energy of the terminal battery $b_{N}$ as $g_1^{(k)} = g_2^{(k)} = \sqrt{g_b \Gamma/2}$ with $\Gamma_a^{(k)}=\Gamma$~\cite{supp}. Figure~\ref{Fig2}(a) presents the example of steady-state energy $E_{2}$ as a function of the interaction phases $\theta_{1}$ and $\theta_{2}$ when choosing the QB number $N=2$ and fixing the interaction strength $g_{b}/\gamma=0.1$. The black-dashed circle illustrates the region of nonreciprocal charging. $E_{2}$ reaches maximum when $\theta_1=\theta_2=-\pi/2$, which can also be verified when extending to cascaded charging of $N$ batteries ($N>2$) by choosing $\theta_k=-\pi/2$, $k\in\{1,2,...,N\}$. Therefore, the optimized steady-state energy of the terminal battery becomes a function of the interaction strengths:
\begin{equation}\label{Ebnr}
    E^{\mathrm{nr}}_{N}/\omega=
    [\frac{2^{2N+1}g_b^{N}\xi}{(2g_b+\gamma)^2(4g_b+\gamma)^{N-1}}]^2.
\end{equation}
Interestingly, nonreciprocity disrupts the pattern of varying charging energy that is dictated by the parity of the number of batteries, indicating a different dynamics in the cascaded charging process~\cite{supp}. 

We take the examples of $N=3$ and $N=4$ to demonstrate the enhancement of nonreciprocal charging by calculating the stored energy of $b_{N}$ in steady state. In addition to the comparison with $E_{N}^{\mathrm{r1}}$ obtained from reciprocal case I, the reciprocal energy ($E_{N}^{\mathrm{r2}}$) in the existence of $a_{k}$ by fixing $\theta_{k}=0$ (denoted as reciprocal case II) is also plotted for comparison. As depicted in Figs.~\ref{Fig2}(b) and (c), we observe that in the weak interaction regime, where the interaction strength is much smaller than the dissipation rate ($g_{b}/\gamma\ll1$), the value of $E_{N}^{\mathrm{nr}}$ is significantly larger than those achieved in the two cases of reciprocal charging, i.e., $E_{N}^{\mathrm{nr}}>E_{N}^{\mathrm{r1}}$ and $E_{N}^{\mathrm{nr}}>E_{N}^{\mathrm{r2}}$. 

The improvement in charging can also be seen from the gain parameters $G_{N1}$ ($G_{N2}$) defined as the ratio of $E^{\mathrm{nr}}_{N}$ and $E^{\mathrm{r1}}_{N}$ ($E^{\mathrm{r2}}_{N}$). For $G_{N1}$, we can neglect the higher-order terms associated with $g_b/\gamma$ in the weak interaction limit, getting $G_{31}\big|_{g_b/\gamma\ll1}\approx [8/(12g_b/\gamma+1)]^2$ and $G_{41}\big|_{g_b/\gamma\ll1}\approx [16/(16g_b/\gamma+1)]^2$. According to Eqs.~\eqref{Ebrn} and~\eqref{Ebnr}, this approximation can also be extended to arbitrary $N$ as $G_{N1}\big|_{g_b/\gamma\ll1}\approx [2^N/(4Ng_b/\gamma+1)]^2$. Apparently, $G_{N1}$ induced by nonreciprocity is significantly enhanced when $g_{b}$ decreases. The upper bound of $G_{N1}$ grows exponentially with the number of batteries: $G_{N1}|_{g_b \rightarrow  0}= 2^{2N}$~\cite{supp}. 
Similarly, we can also derive the gain parameters with respect to reciprocal case II. $G_{N1}>G_{N2}$ can be found in the weak interaction regime of $g_b/\gamma<0.12$, as also demonstrated in Figs.~\ref{Fig2}(d) and (e). Although reciprocal charging is enhanced via the indirect channel implemented by the intermediate modes when $g_b/\gamma<0.1$ ($E_{N}^{\mathrm{r1}}<E_{N}^{\mathrm{r2}}$, see insets), nonreciprocal charging still outperforms: $G_{N1}>G_{N2}\geq 2^{N}$~\cite{supp}.  
This charging gain brought about by nonreciprocity under weak interaction is beneficial, as it allows a substantial enhancement of the charging performance without the need to increase the interaction strength between the quantum systems. Moreover, the exponentially increasing charging gain with the number of QBs also demonstrates the remarkable advantage of nonreciprocity in the cascaded charging of multiple batteries.

In addition to enhanced charging in the weak interaction regime, the maximum value of $E_{N}^{\mathrm{nr}}$ is still larger than $E_{N}^{\mathrm{r1}}$ when the interaction strength is optimized to $g_b^{\mathrm{opt}}=[N+\sqrt{N(8+N)}]\gamma/8$ in the case where $N$ is an odd number. The difference between the maximum value $E_{\max}^{\mathrm{nr}}$ and $E_{\mathrm{max}}^{\mathrm{r1}}$ becomes larger when increasing the number of QBs and the ratio can be fitted as $E^{\mathrm{nr}}_{\max}/E^{\mathrm{r1}}_{\max}\approx1+0.062\ln N$[Fig.~\ref{Fig2}(f)]. It indicates that although the optimal nonreciprocal charging performs identically to reciprocal charging at $N=1$, its superiority in energy charging becomes increasingly pronounced and advantageous as the number of QBs grows in cascaded setup with odd-number QBs.

\begin{figure}
    \includegraphics[angle=0,width=\linewidth]{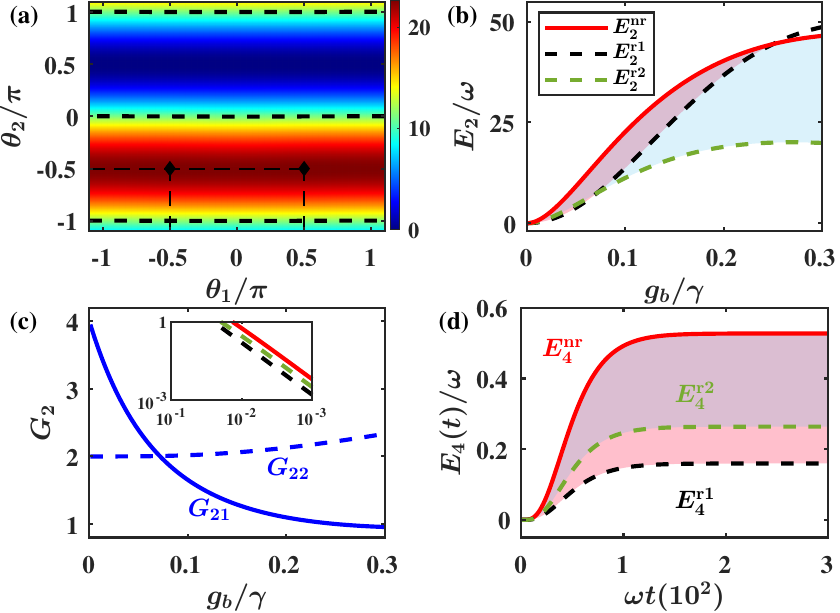}
    \caption{(a) $E_{2}/\omega$ as the function of  $\theta_1$ and $\theta_2$ under $g_b/\gamma=0.1$ when $N=2$. (b) Steady-state $E_2^{\mathrm{nr}}/\omega$, $E_2^{\mathrm{r1}}/\omega$ and $E_2^{\mathrm{r2}}/\omega$ as functions of $g_b/\gamma$ when $N=2$. (c) The gain factor $G_{21}$  and $G_{22}$  as functions of $g_b/\gamma$ when $N=2$. The inset plots the steady-state energy in three cases. (d) The dynamical evolution of $E^{\mathrm{nr}}_4(t)$, $E^{\mathrm{r1}}_4(t)$ and $E^{\mathrm{r2}}_4(t)$ with $g_b=\gamma/100$ when $N=4$. The left parameters are set to be identical to those in Fig.~\ref{Fig2}.}
    \label{Fig3}
\end{figure}

\textit{Parallel nonreciprocal charging.---}
As shown in Fig.~\ref{Model}(c), another typical multi-battery system is the parallel charging setup that all the QBs are charged simultaneously by the common QC. Similarly to the cascaded model, we can derive the Hamiltonian of the parallel charging setup in the rotating frame as $H_{\mathrm{p}}=\sum_{k = 1}^N (g_b^{(k)} e^{i \theta_k} c b_k^\dagger+g_1^{(k)} c a^\dagger_k  + g_2^{(k)} a_k b_k^\dagger)+ \xi c+ H.c.,$ 
where all modes are also assumed to be resonant with respect to the driving field.
In the reciprocal case I without $a_{k}$,  the stored energy of $b_{k}$ in steady state can be obtained as~\cite{supp}:
\begin{equation}\label{rEbpar}
    \begin{split}
    E_k^{\mathrm{r1}}/ \omega=\frac{16 \xi^2 g^2_b (\{\gamma\}/\gamma_b^{(k)})^2}{(4g_b^2\sum_{k = 1}^{N}(\{\gamma\}/\gamma_b^{(k)}) +\{\gamma\})^2},
    \end{split}
\end{equation}
where $ \{ \gamma\}  =\gamma_c \prod_{k = 1}^{N}\gamma_b^{(k)} $. By adding the intermediate modes $a_{k}$ between $c$ and $b_{k}$, the interaction phases $\theta_{k}$ can also be tuned to satisfy and optimize the nonreciprocal charging.
Figure~\ref{Fig3}(a) shows the example of steady-state energy $E_{2}$ for $N=2$ as a function of $\theta_{1}$ and $\theta_{2}$, where the maximum of $E_{2}$ can be found at $\theta_2 = -\pi/2$ and $\theta_1 = \pm \pi/2$. Different from the cascaded charging that we aim to maximize the energy of the terminal battery in the chain, all batteries in the parallel charging can be optimized at the same time. 
Extending to the parallel charging of $N$ batteries ($N>2$), $\theta_k=-\pi/2$ can be derived as the optimal phases for nonreciprocal parallel charging. Therefore, the energy of $b_{k}$ in steady state by plugging into $\theta_{k}=-\pi/2$ is given as:
\begin{equation}\label{nrEbpar}
    \begin{split}
    E_k^{\mathrm{nr}} /\omega=\frac{64 \xi^2 g^2_b}{(2g_b+\gamma_b^{(k)})^2 (2N g_b+\gamma_c)^2}.
    \end{split}
\end{equation}
Compared to the battery energy charged in reciprocal case I, the nonreciprocal charging can avoid the dissipation of each battery (i.e., $\gamma_b^{(k)}$) interfering with each other, as can be seen from the denominators in Eqs.~\eqref{rEbpar} and~\eqref{nrEbpar}, respectively. When adjusted $\gamma_b^{(k)}$, the stored energies of the left batteries are not affected, indicating that nonreciprocity can be used to effectively achieve the independent charging of each port in parallel setups.

Assuming $\gamma_b^{(k)}=\gamma$ for simplicity, the energies stored in all QBs are the same as those in the reciprocal (nonreciprocal) case, i.e., $E_k^{\mathrm{r1}}=E_N^{\mathrm{r1}}$ and $E_k^{\mathrm{nr}}=E_N^{\mathrm{nr}}$, from Eqs.~\eqref{rEbpar} and~\eqref{nrEbpar}.
Taking $N = 2$ as an example to compare the performance of  reciprocal and nonreciprocal charging [Fig.~\ref{Fig3}(b)], we can still find the substantial improvement of stored energy brought about by nonreciprocity in weak interaction regime. Here, the charging performance in reciprocal case II is also plotted for comparison. 
Figure~\ref{Fig3}(c) shows the gain parameters $G_{N1}$ ($G_{N2}$) defined as the ratio of $E^{\mathrm{nr}}_{N}$ and $E^{\mathrm{r1}}_{N}$ ($E^{\mathrm{r2}}_{N}$) for parallel charging. In the limit of weak interaction, we can use the higher-order approximation on $G_{N1}$ and find that $G_{N1}\big|_{g_b/\gamma\ll1}\approx [2/((2N+2)g_b/\gamma+1)]^2$. An improvement by up to a gain of 4 can be found when compared to the reciprocal case I. A similar process can also be applied to the reciprocal case II and the upper bound of the charging gain is derived to be $G_{N2}|_{g_b \rightarrow  0}= 2$~\cite{supp}. Different from the cascaded setups, all batteries in parallel systems share the same charging gain, regardless of the value of $N$.
Nonreciprocal charging can also significantly improve the charging speed in the weak interaction regime. As shown in Fig.~\ref{Fig3}(d), the dynamical increase of stored energy $E_{4}(t)$ in nonreciprocal charging is much faster when compared to the reciprocal case.

\begin{figure}
    \includegraphics[angle=0,width=\linewidth]{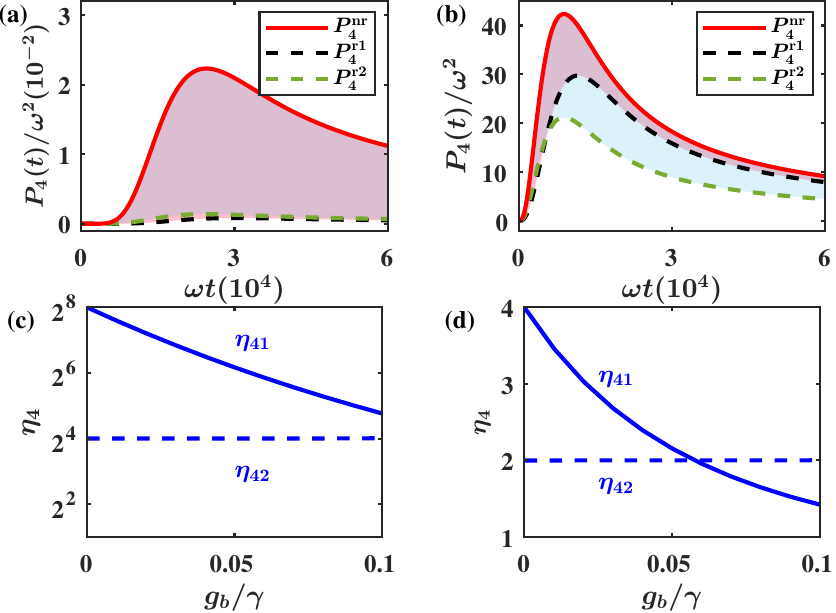}
    \caption{The charging power of terminal battery in cascaded system (a) and the charging power of all batteries in parallel system (b) as functions of $\omega t$ when $N=4$, $g_b=\gamma/10$ and $\xi=\omega$. The ratio ($\eta_{41}$ and $\eta_{42}$) in cascaded system (c) and parallel system (d) as functions of $g_b/\gamma$. We set $\gamma_c=\gamma_b^{(k)} = 0.0005\omega$, $\Gamma=\omega$ and $g_1^{(k)}=g_2^{(k)}=\sqrt{g_b\Gamma/2}$.}
    \label{Fig4}
\end{figure}

\textit{Charging power.---}
Another crucial quantity for assessing the performance of QBs is the charging power $P$, which is defined as $P(t)=E(t)/t$~\cite{PhysRevResearch.2.023113,PhysRevLett.118.150601}, representing the rate at which energy is transferred to the QB during the charging time instance ($t$). 
As shown in Figs.~\ref{Fig4}(a) and (b), we use $N=4$ as an example to compare the reciprocal and nonreciprocal charging power, i.e., $P_{4}^{\mathrm{r1}}$ ($P_{4}^{\mathrm{r2}}$) and $P_{4}^{\mathrm{nr}}$, in cascaded and parallel systems, respectively. 
Significant improvement of the charging power can also be achieved by nonreciprocity when compared to the reciprocal cases in the weak couping regime ($g_{b}/\gamma=0.1$ is set for the plots). 

Since the stored energy of the QB reaches its maximum in steady state, the charging power approaches $0$ in the long-time limit. The maximization of the charging power $P_{\max}$ is obtained by optimizing $t$ in the short-time region.  
To compare the optimal performance of the charging power in nonreciprocal and reciprocal cases, we define the gain parameters as $\eta_{41}=P^{\mathrm{nr}}_{4,\max}/P^{\mathrm{r1}}_{4,\max}$ and $\eta_{42}=P^{\mathrm{nr}}_{4,\max}/P^{\mathrm{r2}}_{4,\max}$. 
As demonstrated in Figs.~\ref{Fig4}(c) and (d), nonreciprocity still leads to an improvement in the maximum value of the charging power for both cascaded (c) and parallel (d) setups, i.e., $\eta_{41}(\eta_{42})>1$, in the weak interaction regime ($g_b/\gamma\in[0,0.1]$). 
In contrast to reciprocal case I, where $\eta_{41}$ increases to its maximum ($\eta_{41,\max}=2^8$ for cascaded charging and $\eta_{41,\max}=4$ for parallel charging) as the interaction strength approaches 0, the change of $\eta_{42}$ within the range of $g_b/\gamma\in[0,0.1]$ is negligible. $\eta_{42}$ is approximately equal to $2^4$ for cascaded charging and $2$ for parallel charging. Extending the calculation to cascaded charging of $N$ batteries, we can find that the upper bound of $\eta_{N1}$ ($\eta_{N2}$) is $2^{2N}$ ($2^{N}$), which is exactly the same as the maximum of the gain parameters $G_{N1}$ ($G_{N2}$) for the stationary energy of $E_{N}^{\mathrm{r1}}$ ($E_{N}^{\mathrm{r2}}$)~\cite{supp}. Consistent with the cascaded case, the same improvement by up to a gain of 4 (2) can be found for $\eta_{N1}$ ($\eta_{N2}$) and $G_{N1}$ ($G_{N2}$) when compared to the reciprocal case I (II) in parallel charging. It shows that nonreciprocity has the same enhancement on both energy storage and charging power in multi-battery systems.

\textit{Conclusion.---} 
In conclusion, by introducing lossy intermediate modes to construct non-Hermitian AB triangles within cascaded and parallel structures, nonreciprocal charging in multi-battery systems can be efficiently achieved. To quantify the enhanced charging performance, we calculate the stored energy and the charging power of the QBs. In the weak interaction regime, when compared with reciprocal scenarios, nonreciprocity brings about the same gain factor for these two quantities.
Notably, as the charging distance increases, cascaded charging exhibits an exponentially growing gain. Meanwhile, nonreciprocity can boost parallel charging up to four times.
Moreover, in a cascaded setup with an odd number of QBs, the maximized stored energy also shows an increasing improvement as the number of QBs increases, highlighting the unique advantage of our scheme in multi-battery charging.

Our protocol offers an efficient approach to overcome the challenge of weak interaction between quantum systems. It holds potential for realizing high-performance QBs and controlling the directional energy flow in quantum networks for applications~\cite{PhysRevLett.120.060601}.    
This scheme can be implemented in a variety of systems, including 
ultracold atoms~\cite{PhysRevLett.124.070402}, optomechanical setup~\cite{PhysRevLett.126.123603} and cavity magnonics~\cite{PhysRevLett.123.127202,magnonics}. 
Considering the strong link between non-Hermitian AB structures and topological phases, further exploration of the impacts of non-Hermitian topology on energy charging could provide valuable insights into a deeper understanding of quantum thermodynamics~\cite{Vinjanampathy_2016,PhysRevLett.120.060601} in a topological framework.

\textit{Acknowledgements.---}
This work is supported by the National Natural Science Foundation of China (Grants No.~12474353, No.~12474354), the Aviation Science Foundation of China (Grants No.~20240058051004) and the Fundamental Research Funds for the Central Universities.

\bibliography{Ref}

\end{document}